\documentclass{aastex63}
\usepackage{hyperref}
\usepackage[colorinlistoftodos]{todonotes}

\shorttitle{Fe K$\alpha$ Centroid in Supernova Remnants}

\begin{document}
\title{Can the Fe K-alpha line reliably predict supernova remnant progenitors?}

\correspondingauthor{Vikram V. Dwarkadas}
\email{vikram@astro.uchicago.edu}

\author[0000-0002-9337-0902]{Jared Siegel}
\affiliation{Department of Astronomy and Astrophysics, University of Chicago \\
5640 S Ellis Ave \\
Chicago, IL 60637, USA}
\affil{Center for Interdisciplinary Exploration and Research in Astrophysics, Northwestern University \\
1800 Sherman Ave \\
Evanston, IL 60201, USA}

\author[0000-0002-4661-7001]{Vikram Dwarkadas}
\affiliation{Department of Astronomy and Astrophysics, University of Chicago \\
5640 S Ellis Ave \\
Chicago, IL 60637, USA}

\author[0000-0003-0570-9951]{Kari A. Frank}
\affil{Center for Interdisciplinary Exploration and Research in Astrophysics, Northwestern University \\
1800 Sherman Ave \\
Evanston, IL 60201, USA}

\author[0000-0003-0729-1632]{David N. Burrows}
\affiliation{Department of Astronomy and Astrophysics, Pennsylvania State University \\
525 Davey Laboratory \\
University Park, PA 16802, USA}

\begin{abstract}

The centroid energy of the Fe K$\alpha$ line has been used to identify the progenitors of supernova remnants (SNRs). These investigations generally considered the energy of the centroid derived from the spectrum of the entire remnant. Here we use {\it XMM-Newton} data to investigate the Fe K$\alpha$ centroid in 6 SNRs: 3C~397, N132D, W49B, DEM L71, 1E 0102.2-7219, and Kes 73. In Kes 73 and 1E 0102.2-7219, we fail to detect any Fe K$\alpha$ emission. We report a tentative first detection of Fe K$\alpha$ emission in SNR DEM L71, with a centroid energy consistent with its Type Ia designation. In the remaining remnants, the spatial and spectral sensitivity is sufficient to investigate spatial variations of the Fe K$\alpha$ centroid. We find in N132D and W49B that the centroids in different regions are consistent with that derived from the overall spectrum, although not necessarily with the remnant type identified via other means. However, in SNR 3C~397, we find statistically significant variation in the centroid of up to 100 eV, aligning with the variation in the density structure around the remnant. These variations span the intermediate space between centroid energies signifying core-collapse and  Type Ia remnants. Shifting the dividing line downwards by 50~eV can place all the centroids in the CC region, but contradicts the remnant type obtained via other means. Our results show that caution must be used when employing the Fe K$\alpha$ centroid of the entire remnant as the sole diagnostic for typing a remnant.

\end{abstract}

\section{Introduction}
\label{sec:intro}

Supernova (SN) explosions result from the core-collapse (CC) and explosion of a massive star, or the thermonuclear deflagration and detonation of a white dwarf in a binary system (Type Ia). The resulting explosion drives a collisionless shock into the surrounding medium, which sweeps up the material and expands over hundreds of years to form large structures of gas and dust. The explosion can release the products of stellar and supernova nucleosynthesis into the ambient medium. The SN shock itself heats up the surrounding medium, causing it to emit across the entire wavelength range. Thus supernova remnants (SNRs) can provide unique insights into both the SN explosion itself, the ejected SN material, and the nature of the surrounding medium. However, isolating the contributions of the products of the SN explosion, the progenitor mass-loss, and the surrounding medium presents a considerable challenge to understanding the origin and evolution of SNRs. While observations of SNRs can shed light on their abundance distribution, morphology, dynamics, and kinematics, there is generally no simple way to relate these features to the SN progenitor. Sometimes, observation of a central compact object can directly point to a core-collapse origin. In other cases, the abundance distribution is a giveaway of a Type Ia progenitor. However, in many intermediate cases, the observations do not provide a clear-cut view of a remnant's origins.

Past studies have attempted to investigate SNR types through their X-ray emission. For instance, in young SNRs the observed abundance pattern can reveal signatures of the explosion mechanism \citep{Hughes1995, Vink2012}. Alternatively, \citet{Lopez2011} argued that Type Ia SNRs are more symmetric than core-collapse SNRs, using the power-ratio method; however, there are notable exceptions, such as SNR 1E 0102-7219, a highly symmetric CC remnant \citep{Hughes2000, Eriksen2001}; RCW 86, a Type Ia SNR that may have evolved in a wind bubble \citep{williamsetal11}; and N103B, a Type Ia SN which may be expanding into an hourglass-shape cavity, forming  bipolar bubbles of ejecta \citep{yamaguchietal21}.

\citet[][hereafter Y14]{yamaguchi2014} proposed a method to identify the SNR type based on the Fe K$\alpha$ emission line. Using the complete sample of SNRs observed with \textit{Suzaku}, Y14 extracted the spectra from the entire emitting region of each remnant (with the exception of IC 443, due to its large angular size) and fit the emission within the 5-10 keV band with an absorbed powerlaw and Gaussian model, with additional Gaussian or radiative recombination continuum (RCC) components added as needed. They concluded that all SNRs confidently classified as Type Ia had an Fe K$\alpha$ centroid energy less than 6550 eV, while SNRs confidently classified as CC had energies greater than 6550~eV, and clearly stated that no single object with a  robust progenitor fell on the wrong side of this dividing line. Y14 also observed that within each progenitor type (Ia or CC), the remnants' Fe K$\alpha$ centroid energy and luminosity are potentially correlated, such that those which had a higher level of Fe ionization tended to have a more luminous Fe K$\alpha$ line. They postulated that the centroid energy of the Fe K$\alpha$ line could be used to identify a remnant's type (Ia or CC) and treated 6550~eV as the dividing line. Such a typing method for SNRs would not depend on uncertain or difficult-to-measure quantities such as distance. Since the energy of the centroid depends on the ionization state of Fe, determined by the history of the shocked Fe, Y14 attributed the distinction between the centroid energy in Type Ia and CC SNRs to the lower density ambient media typically found around Type Ia SNe. However, Kepler's SNR, where a Type Ia remnant shows evidence of circumstellar interaction, may be a counterexample to this \citep{Chiotellis2012,Patnaude2012,Chiotellis2020}. 

Y14 found that results from the \textit{Suzaku} sample were consistent with theoretical Type Ia models evolving into uniform ambient media. Further modeling extended this agreement to CC remnants as well, with the exception of some high luminosity SNR, including N132D and W49B \citep{patnaudeetal15}. 1D theoretical modeling, followed by a calculation of the X-ray line emission, predicted a centroid energy-luminosity correlation for each progenitor type, with an overlap of centroid energies between CC and Type Ia remnants near 6000~eV \citep{patnaudeetal15,Martinez-Rodriguez2018,Jacovich2021}. Near this energy, the models predicted that luminosity would be the primary discriminator.

While the centroid energy-luminosity correlation can be seen in both theoretical modeling and Y14, many recent studies \citep{sezeretal18, bhaleraoetal19, sawadaetal19, quirolavasquezetal19,Martinez-Rodriguez2020} have continued to treat the Fe K$\alpha$ line centroid as the sole discriminant. 

\begin{deluxetable*}{lcccccccc}[t]
\tablecaption{\textit{XMM-Newton} Observations\label{tab:obs}}
\tablehead{
\colhead{Name} & \colhead{EPIC ID} &  \colhead{pn} &
\colhead{MOS1} & \colhead{MOS2} & \colhead{$N_H$} & \colhead{Distance} & \colhead{Age} & \colhead{References}\\
& &(ks)&(ks)&(ks)&($10^{22}$ cm$^{-2}$)&(kpc)&(yrs)
}
\startdata
3C~397 & 0085200401 & 12.6 & 15.7 & 15.8 & 3 & 8 & 1350-5300 & 1, 2\\
N132D & 0210681301 & 18.4 & 15.0 & 15.0 & 0.06 & 50 & 3150 & 3\\
W49B & 0724270101 & 115.5 & 117.2 & 117.1 & 5 & 8 & 1000-6000 & 4, 5, 6\\
DEM L71 & 0201840101 & 60.3 & 62.0 & 62.0 & 0.07 & 50 & 4400 & 7\\
1E 0102.2-7219 & 0135721701 & 25.3 & \textbf{---} & \textbf{---} & 0.05 & 60 & 1000-2100 & 8, 9\\
Kes 73 & 0013340201 & \textbf{---} & 6.4 & 6.4 & 3 & 8.5 & 750-2100 & 10, 11\\
\enddata
\tablecomments{(1): \cite{SafiHarb2005}; (2): \cite{Leahy2016}; (3): \cite{Morse1995}; (4): \cite{pye1984}; (5): \cite{Zhu2014}; (6): \cite{zv18}; (7): \cite{Ghavamian2003}
; (8): \cite{Hughes2000}; (9): \cite{Eriksen2001}; (10): \cite{Tian2008}; (11): \cite{Kumar2014}.}
\end{deluxetable*}

\begin{figure*}[ht]
\gridline{\fig{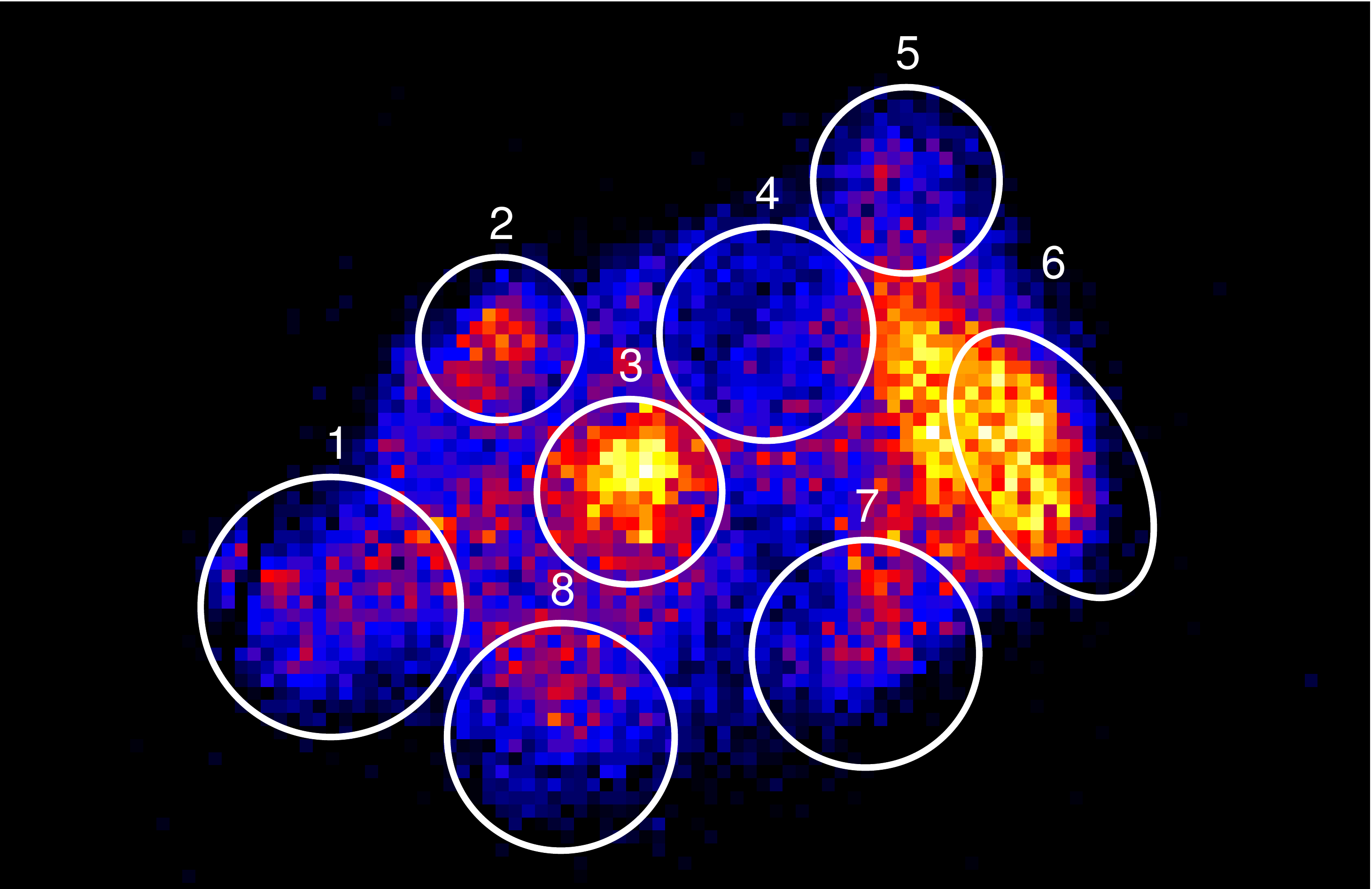}{0.33\textwidth}{} \fig{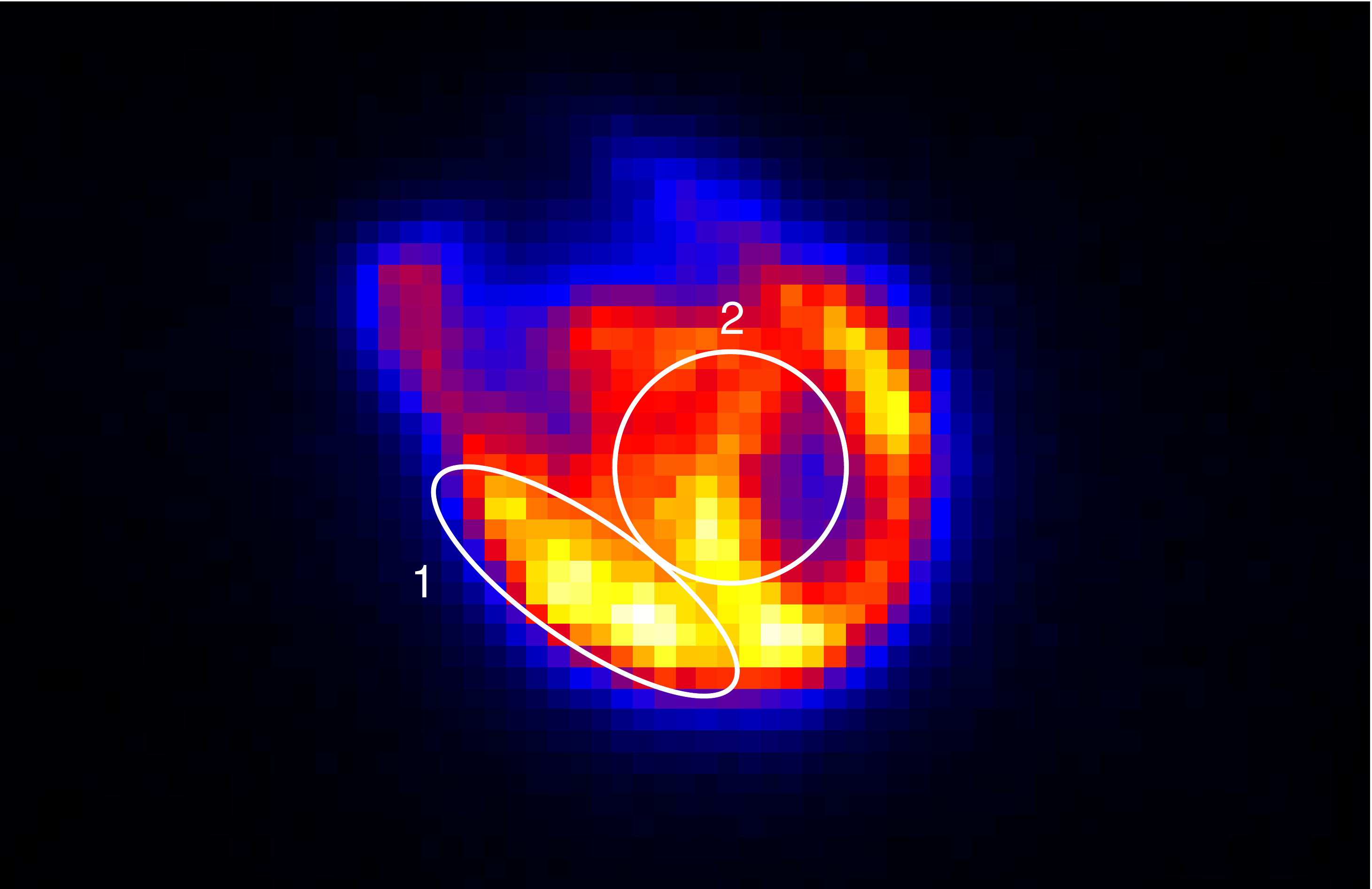}{0.33\textwidth}{} \fig{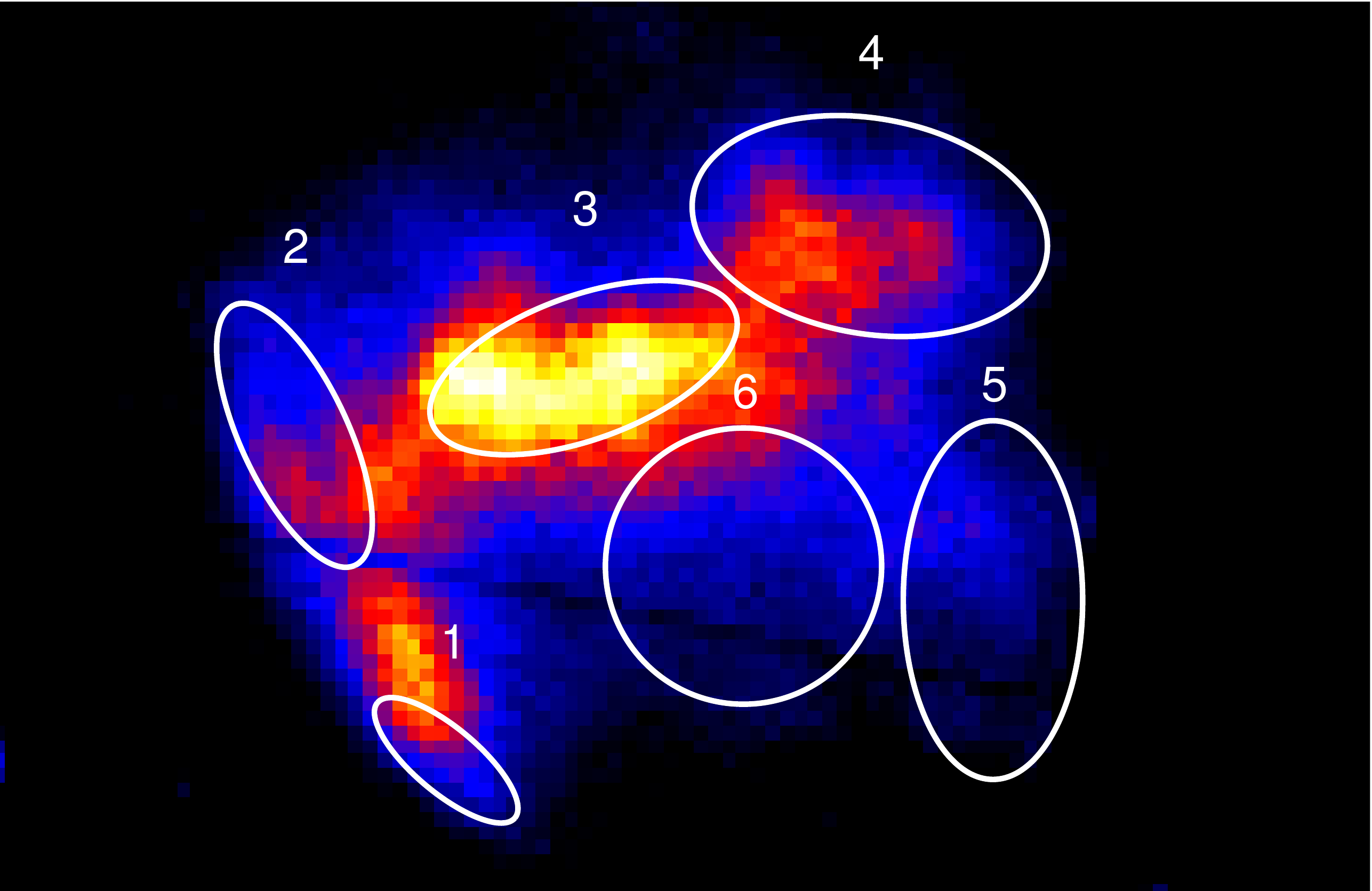}{0.33\textwidth}{}}

\caption{From left to right: EPIC pn images of 3C~397, N132D, and W49B. The numbered regions correspond to the regions for our spatially resolved Fe K$\alpha$ centroid fits.}
\label{fig:obs}
\end{figure*}

The Y14 calibration appears to hold for many SNRs. However, the robustness of this method has not been critically examined. Recently, \citet{siegeletal20b} suggested that although the Fe K$\alpha$ centroid energy for the SNR W49B falls in the core-collapse regime, the elemental abundances suggest that the remnant may in fact have a Type Ia origin, as pointed out earlier by \cite{zv18}. \citet{siegeletal20b} postulated that the higher centroid energy for W49B resulted from the remnant's expansion into high density media. In addition to W49B, a growing number of Type Ia candidate remnants have been suggested to be expanding into high density media, including 3C~397 \citep{SafiHarb2005, Leahy2016} and N103B \citep{vanderHeyden2002, Rest2005}. Sgr A East presents another interesting case. A recent abundance study favored a Type Iax origin \citep{Zhou2021}, but the Fe K$\alpha$ centroid energy reported by Y14 places it in the core-collapse SNR category.

Expansion of the SNR shock wave into a high density medium would result in a reflected shock expanding back into the SNR \citep{cl89,tbfr90,trfb91,Dwarkadas2005,dwarkadas07}, which could ionize the Fe and shift the centroid to higher energies. Alternatively, Type Ia remnants expanding into a high density from the time of explosion would also show a higher centroid energy. For instance, \cite{Bochenek2018}  found that SN 2012ca was interacting with a region of density $> 10^6$ cm$^{-3}$, and perhaps as high as $10^8$ cm$^{-3}$, suggesting a high density medium at a distance of about 5 $\times 10^{15}$ cm from the SN explosion. \citet{alderingetal06} suggested that the circumstellar medium around SN 2005gj had a density as high as 10$^{10}$ cm$^{-3}$. While the source of the high density material is not apparent, the entire subset of Type Ia-CSM SNe appear to show interaction of the SN shock wave with a high density medium close in to the SN \citep{silvermanetal13}. These factors would move the Fe K$\alpha$ centroid to higher energies.

Many SNRs are large, parsec-scale objects, and the density around them varies, often considerably. The large size means that portions of the shock front may sometimes expand into clouds, clumps or other dense material. If the centroid of the Fe K$\alpha$ line depends on the density into which a remnant expands, then the centroid energies across a remnant's surface may not be homogeneous, and in certain regions the energies may shift above or below 6550~eV. Theoretical 1D models \citep{patnaudeetal15,Martinez-Rodriguez2018,Jacovich2021}, have generally confirmed the correlation between ambient density and the centroid energy of the entire remnant. However, point-to-point variations of the Fe K$\alpha$ centroid energy within a single SNR have neither been predicted nor reported by these authors. The spherically symmetric models, by their very nature, are unable to model the Fe K$\alpha$ centroid energy variations that can arise due to density inhomogeneities around a single SNR. The latter is the main consideration in this paper.

Motivated by these factors, in this paper we investigate the robustness of the Fe K$\alpha$ centroid energy as a diagnostic of SNR type through spatially resolved spectral fits. In \S \ref{sec:obs} we discuss the observational data used in this paper and the fits to the Fe K$\alpha$ line, as well as report the line centroids and photon luminosities. We also discuss the Fe K$\alpha$ line in the SNR DEM L71 in significant detail. \S \ref{sec:disc} discusses the results, especially for the spatially resolved remnants. Conclusions are summarized in \S \ref{sec:concl}.

\section{Observations and Data Analysis}
\label{sec:obs}
We analyze archival \textit{XMM-Newton} data for 6 SNRs (listed in Table \ref{tab:obs}). The remnants were selected to ensure significant levels of measurable thermal emission, sizes larger than the \textit{XMM-Newton} angular-resolution, and observations with $\gtrapprox 10^5$ photons. In addition to these criteria, we also sought remnants that previous studies of SNR Fe K$\alpha$ emission identified as outliers, such as 3C~397 and W49B \citep{yamaguchi2014, patnaudeetal15, siegeletal20b}.

\begin{deluxetable}{lccccccc}[t]
\tablecaption{Spectral Fitting Results\label{tab:res}}
\tablehead{
\colhead{Remnant} & \colhead{Region} &  \colhead{Fe K$\alpha$ Centroid Energy} & \colhead{Photon flux} & \colhead{Photon luminosity}\\
 &  & (keV) & ($10^{-5}$ cm$^2$ s$^{-1}$) & ($10^{40}$ s$^{-1}$)
}
\startdata
3C 397 & Overall & $6.578^{+0.007}_{-0.006}$ & $14.6^{+1.0}_{-1.1}$ & $112^{+7}_{-8}\rm \times d_{8}^2$\\
 & 1 & $6.53^{+0.02}_{-0.02}$ & $1.5^{+0.3}_{-0.3}$ & $11^{+2}_{-2}\rm \times d_{8}^2$\\
 & 2 & $6.63^{+0.03}_{-0.03}$ & $0.5^{+0.2}_{-0.2}$ & $4^{+1}_{-1}\rm \times d_{8}^2$\\
 & 3 & $6.55^{+0.02}_{-0.02}$ & $1.3^{+0.2}_{-0.2}$ & $10^{+2}_{-2}\rm \times d_{8}^2$\\
 & 4 & $6.58^{+0.03}_{-0.03}$ & $0.6^{+0.2}_{-0.2}$ & $5^{+1}_{-1}\rm \times d_{8}^2$\\
 & 5 & $6.61^{+0.03}_{-0.03}$ & $0.9^{+0.2}_{-0.2}$ & $7^{+2}_{-2}\rm \times d_{8}^2$\\
 & 6 & $6.60^{+0.02}_{-0.01}$ & $1.4^{+0.2}_{-0.2}$ & $11^{+2}_{-2}\rm \times d_{8}^2$\\
 & 7 & $6.62^{+0.04}_{-0.04}$ & $0.7^{+0.2}_{-0.2}$ & $5^{+2}_{-1}\rm \times d_{8}^2$\\
 & 8 & $6.56^{+0.04}_{-0.03}$ & $0.8^{+0.2}_{-0.2}$ & $6^{+2}_{-2}\rm \times d_{8}^2$\\
\hline
N132D & Overall & $6.69^{+0.02}_{-0.02}$ & $1.3^{+0.3}_{-0.3}$ & $400^{+80}_{-80}\rm \times d_{50}^2$\\
 & 1 & $6.62^{+0.04}_{-0.04}$ & $0.3^{+0.2}_{-0.1}$ & $80^{+50}_{-30}\rm \times d_{50}^2$\\
 & 2 & $6.67^{+0.04}_{-0.04}$ & $0.3^{+0.1}_{-0.1}$ & $100^{+40}_{-30}\rm \times d_{50}^2$\\
\hline
W49B & Overall & $6.6647^{+0.0006}_{-0.001}$ & $102.5^{+0.8}_{-0.8}$ & $785^{+6}_{-6}\rm \times d_{8}^2$\\
 & 1 & $6.669^{+0.005}_{-0.006}$ & $2.0^{+0.1}_{-0.1}$ & $15.0^{+0.8}_{-0.8}\rm \times d_{8}^2$\\
 & 2 & $6.660^{+0.003}_{-0.002}$ & $6.9^{+0.2}_{-0.2}$ & $53^{+1}_{-1}\rm \times d_{8}^2$\\
 & 3 & $6.660^{+0.001}_{-0.002}$ & $21.4^{+0.4}_{-0.3}$ & $164^{+3}_{-3}\rm \times d_{8}^2$\\
 & 4 & $6.660^{+0.011}_{-0.003}$ & $7.1^{+0.2}_{-0.2}$ & $55^{+2}_{-1}\rm \times d_{8}^2$\\
 & 5 & $6.645^{+0.015}_{-0.002}$ & $2.3^{+0.1}_{-0.1}$ & $17.7^{+1}_{-0.9}\rm \times d_{8}^2$\\
 & 6 & $6.660^{+0.005}_{-0.003}$ & $5.8^{+0.2}_{-0.2}$ & $45^{+1}_{-1}\rm \times d_{8}^2$\\
\hline
DEM L71 & Overall & $6.45^{+0.05}_{-0.05}$ & $0.1^{+0.05}_{-0.04}$ & $30^{+10}_{-10}\rm \times d_{50}^2$\\
\enddata
\tablecomments{For photon luminosity, we adopt the distances of Table \ref{tab:obs} and introduce the scaled distance d$_{X}$ = d/($X$~kpc). }
\end{deluxetable}

\subsection{Iron K-Shell Line Fitting}
\label{sec:fitting}

For each remnant, the data were reduced using XMM SAS 18.0.0. Calibrated event lists were created using the \texttt{epproc} task. Flares due to high particle background were removed following the standard filtering procedures as described in SAS documentation\footnote{https://www.cosmos.esa.int/web/xmm-newton/sas-thread-epic-filterbackground}. Locations with known background point sources from the literature were removed from the calibrated and filtered event files, as were additional background point sources detected with \texttt{wavdetect}\footnote{https://cxc.harvard.edu/ciao/ahelp/wavdetect.html}. We extracted spectra for the entire emitting region. We limit our measurements to the pn detector, since the effective area of the pn detector is much higher than that of the two MOS cameras\footnote{https://xmm-tools.cosmos.esa.int/external/xmm\_user\_support/documentation/uhb/effareaonaxis.html}. Since the angular sizes of the remnants are considerably smaller than the field-of-view, background spectra are extracted from regions surrounding the SNRs and subtracted to account for local and instrumental background.

We isolate the spectra from approximately 5 to 10 keV and model the emission with an absorbed powerlaw continuum and a Gaussian component for the Fe K$\alpha$ line; for 3C~397 and W49B, additional Gaussian components are added for Cr, Mn, and Ni emission. To facilitate comparison, our spectral fitting procedure is intended to closely parallel Y14. We find that the resulting centroid measurements are quite robust and not dependent on the details of the model; adding or removing additional Gaussian components (where relevant), changing the selected energy range, using a Tuebingen-Boulder ISM absorption model instead of \texttt{phabs}, or using a thermal bremsstrahlung continuum instead of a power law has negligible impact on the centroid results.

We detect Fe K$\alpha$ emission in four of the six remnants; non-detections are reported for Kes 73 and 1E 0102.2-7219. We note that the Fe K$\alpha$ line was not detected in DEM L71 by Y14, or reported by other authors, but is tentatively detected in our study (see \S \ref{sec:DEML71} for discussion).

\begin{figure*}[t]
\gridline{\fig{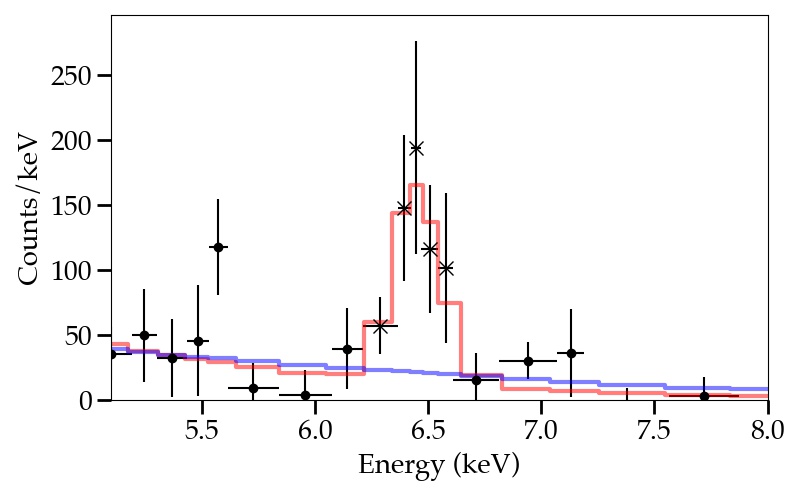}{.7\textwidth}{}}

\caption{ SNR DEM L71 spectrum from $5-8$~keV (black points), along with the best-fit absorbed powerlaw model (blue curve), and the best-fit absorbed powerlaw-Gaussian model (red curve). The five bins associated with the emission line for the \cite{Kraft1991} significance test are plotted as crosses (see Section \ref{sec:DEML71} for discussion).}

\label{fig:DEML71}
\end{figure*}

Y14 categorized the non-detection cases into two groups: low kT evolved SNRs and young SNRs dominated by non-thermal emission. Both Kes 73 \citep{Tian2008, Kumar2014} and 1E 0102.2-7219 \citep{Hughes2000, Eriksen2001, Neslihan2019} fall into the first category.

The presence of overionized plasma merits additional consideration in the case of W49B. \cite{ozawaetal09} first reported the presence of overionized plasma in W49B; Y14 later identified a radiative recombination continuum (RRC) of Fe XXV in the X-ray spectrum. Both added an exponential component with a threshold energy of $\gtrapprox$ 8 keV to fit the RRC. We find that due to the low flux at energies $\gtrapprox$ 8 keV, the inclusion of such an exponential component does not impact our results.

For the remnants N132D, W49B, and 3C~397, we find the Fe K$\alpha$ emission is strong enough and the remnants are large enough in the plane of the sky that we can conduct spatially resolved Fe K$\alpha$ centroid studies. Due to the low count rates for energies $\gtrapprox$ 6 keV, a similar analysis is not possible for the SNR DEM L71. Consequently we investigate the Fe K$\alpha$ line in various regions of N132D, W49B, and 3C~397. The selected regions are shown in Figure \ref{fig:obs}. For each region, we extract the pn spectra and follow the same fitting procedure as that used for the entire remnant.

\begin{figure*}[tbp]
\gridline{\fig{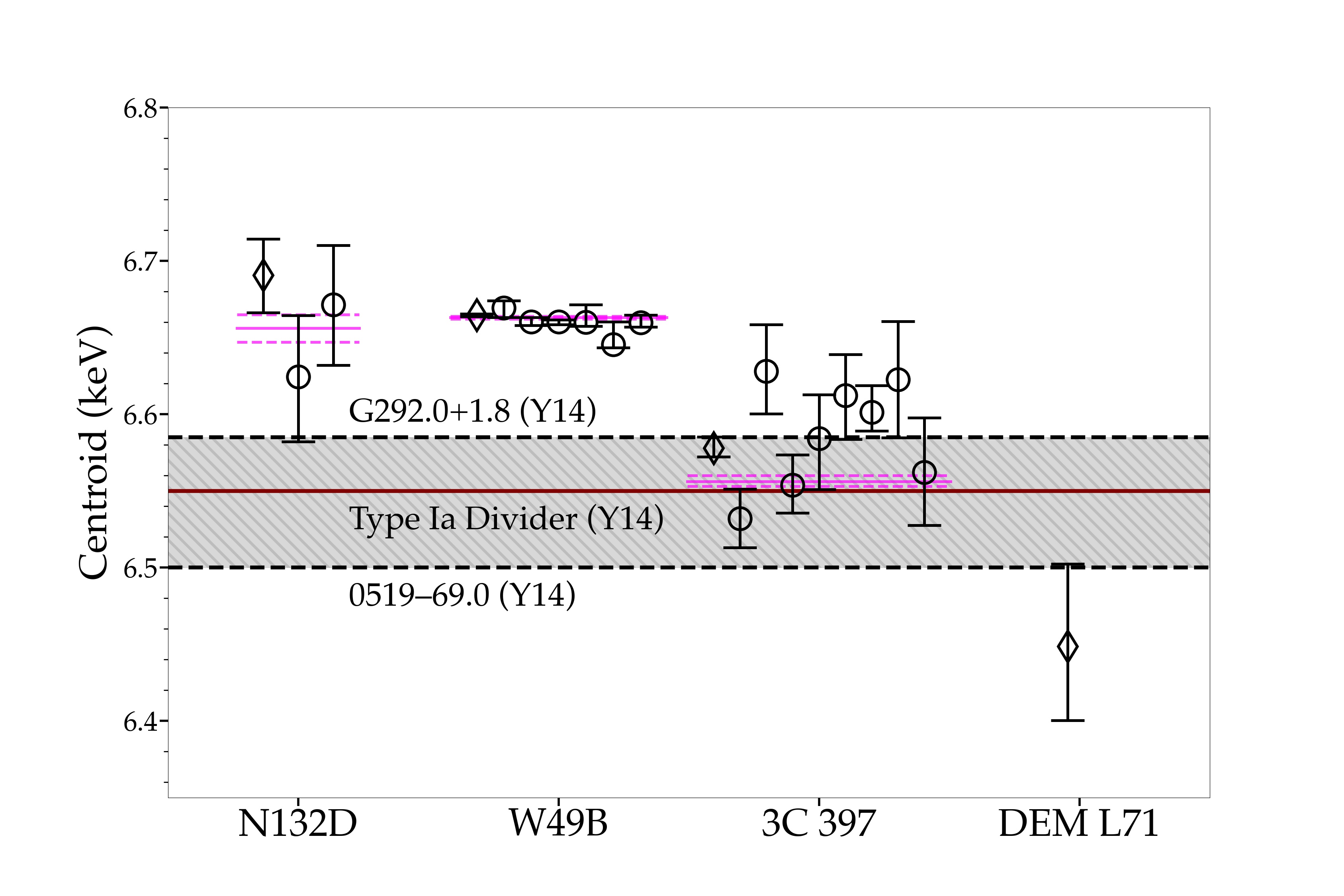}{.7\textwidth}{}}

\caption{A comparison of the Fe K$\alpha$ centroid energies for each remnant, with the results of Y14 shown as magenta lines. For each remnant, the centroid for the entire emitting region is shown as a diamond, with the spatially resolved regions placed to the right. The dashed-black lines represent the remnant with the lowest overall centroid energy of those confidently classified as core-collapse (SNR G292.0+1.8) and the remnant with the highest centroid energy of those confidently classified as Type Ia (SNR 0519–69.0) from Y14. The Type Ia dividing line from Y14 is presented as a solid-red line at 6550~eV.}
\label{fig:centroids}
\end{figure*}

From our Fe K$\alpha$ line fits, we report the centroid energy (keV), photon flux, and photon luminosity of the Fe K$\alpha$ line in each region. These results are presented in Table \ref{tab:res}. The Fe K$\alpha$ centroids for each remnant are compared in Figure \ref{fig:centroids}, along with the data from Y14. The quoted uncertainties are the 90\% confidence levels.

The overall luminosities of the Fe K$\alpha$ line in 3C~397, N132D, and W49B are comparable to those found by Y14, with N132D being about 30\% lower. Some of this difference may arise from the difference between \textit{XMM-Newton} and \textit{Suzaku} response at the energy of the Fe K$\alpha$ line, and perhaps due to differences in the line fitting. The overall photon flux from the Fe K$\alpha$ line in DEM L71 is much lower than that from the other remnants, consistent with the fact that the line was not seen in earlier observations.

\subsection{DEM L71}
\label{sec:DEML71}

As discussed in \S \ref{sec:fitting}, we fit the DEM L71 spectrum above $5$~keV to an absorbed powerlaw with a single Gaussian component; the best-fit model yields a reduced $\chi^2$ of $23.18/18$. For comparison, we also fit the spectrum to an absorbed powerlaw without a Gaussian component, which yields a reduced $\chi^2$ of $41.66/21$. The best fit models for both variants are presented alongside the DEM L71 spectrum in Figure \ref{fig:DEML71}. Prior studies have not reported Fe K$\alpha$ line emission from SNR DEM L71, although the remnant itself has been well studied.  Given the low number of counts within the line, we conduct a series of tests to weigh the significance of our detection.

Following \cite{Bevington1969}, we first use the F-statistic to test for the presence of an additional model component. For two given models, the first of which is nested within the second but does not lie on the boundary of the parameter space of the second model, the F-test quantifies the improvement of the spectral fit due to an additional additive component. Using our absorbed powerlaw model and absorbed powerlaw-Gaussian model, we find an F-statistic of $4.78$, which corresponds to a null-hypothesis probability of $0.013$. This tentatively supports the addition of a Gaussian component at a $95\%$ confidence level. However, it is not clear that the F-test is reasonable for identifying the presence of a line (although it has often been used to do so). The absorbed power-law model without a Gaussian could be considered to be implicitly using a Gaussian with zero intensity, thus lying on the boundary of the model with the additive component, and therefore violating the assumptions necessary for the F-test to be applicable \citep{Protassov2002}. 

Given the caveats with the F-test, we next consider the method of \cite{Kraft1991}. For a given $N$ observed counts and expected background continuum of $N_b$ counts, the number of counts from the source is simply $N_s=N-N_b$. If $N_s$ is non-zero to high confidence, the candidate signal may be considered a source detection. Although $N_s$ and $N_b$ are not directly observable, $N$ is easily accessible. Following \cite{Kraft1991} we thus assume the source and background are both Poisson-distributed and calculate the confidence interval of $N_s$ for a given $N$ and assumed $N_b$. Between $6.2-6.6$~keV, we calculate a total of $N=48$~counts; the included bins are highlighted as crosses in Figure \ref{fig:DEML71}. To determine the background continuum, we first integrate the absorbed-powerlaw model over the same energy range; this yields $9$~counts from the continuum. To reflect the uncertainty in the continuum counts, we calculate the confidence interval for the background counts, again assuming a Poisson-distribution; at a $95\%$ confidence level, this yields $2-12$~counts. Adopting the upper limit for $N_b$, we find a $95\%$ confidence interval for $N_s$ of $24-51$~counts. We thus conclude that the feature near $6.5$~keV in SNR DEM L71 constitutes a line detection with high confidence.

While the identification of the feature as a line appears solid, the association with Fe K$\alpha$ may be tenuous. In addition to Fe K$\alpha$, there are other emission lines which could produce the observed signal near $6.5$~keV, including Mn XXV, Fe XX, and Fe XXI. A clue to the line identification comes from \cite{franketal19}, who found that DEM L71 hosts a reverse-shocked pure-ejecta component, characterized by Fe abundance above solar and $kT>1$~keV. The Fe K$\alpha$ line is a fluorescent line that arises from neutral Fe interacting with high-energy photons $>$ 6.4 keV. In practice photons with energy above 8 keV are generally needed to fluoresce it from a continuum source. Given the high temperature in the shocked interior of DEM L71, and the large amount of Fe, it is plausible to assume that there are sufficient high-energy photons to produce the faint emission.  Therefore we tentatively identify the feature as Fe K$\alpha$. 

\section{Discussion}
\label{sec:disc}

The Fe K$\alpha$ centroid energies derived from the overall spectrum of SNRs 3C~397, N132D, and W49B are in general agreement with those found in Y14 (Figure \ref{fig:centroids}). Our tentative Fe K$\alpha$ detection in DEM L71 falls comfortably within the Type Ia parameter space reported by Y14. This classification is consistent with several previous studies of DEM L71, which identified Fe enhancements and pure-ejecta emission \citep{Ghavamian2003, hughesetal03, franketal19, Siegel2020}.

The spatially resolved fits show minimal spread in N132D and W49B, in line with the energy of the overall centroid. In N132D, we do not detect Fe K$\alpha$ emission along the south-western rim, and thus are limited to interior regions for this remnant; this was also noted by \citet{Sharda2020}. 

In the case of W49B, all the centroid energies lie in the CC parameter space, in agreement with the overall centroid energy. This would consequently result in a CC designation for the remnant. However recent papers have suggested that the remnant could have a Type Ia origin \citep{zv18, siegeletal20b}, based on the abundance distribution and the large Fe content.

The remnant 3C~397 presents an interesting case. We find statistically significant variation in the Fe K$\alpha$ centroid across various regions; the energy of the centroid varies by about 100 eV over the remnant surface. While this variation is large, we note that \citet{Fukushima2020} report a variation of $\sim$ 130 eV for various Fe K$\alpha$ centroid energies in G344.7–0.1. The difference is that in G344.7-0.1, the range of variations was fully confined to the Type Ia region elucidated by Y14. For Kes 79, \citet{sezeretal18} report a variation of 50 eV between different concentric regions, also confined to the Type Ia region. In 3C~397 however, the centroid energies of the different regions fall on either side of the Y14 dividing line, while also overlapping the centroid energy of SNR G292.0+1.8, a CC remnant with a known pulsar (see Figure \ref{fig:centroids}). The regions with the highest centroid energies (labeled 2, 5, 6, and 7) lie along the northern and western shell of 3C~397 (see Table \ref{tab:res} and Figure \ref{sec:obs}). CO surveys have shown that these regions are located along the site of interaction between 3C~397 and a molecular cloud \citep{SafiHarb2005, Kilpatrick2016}. It is possible that the high centroid energies along the northern and western fronts are indicative of shock interaction with the molecular cloud followed by a reflected shock expanding back into the SNR, ionizing the Fe. Meanwhile, regions along the eastern and southern shells (labeled 1, 3, 4, and 8) show lower centroid energies, more consistent with the Type Ia designation. They appear to be coincident with a lower density ambient medium \citep{SafiHarb2005}. 

The spatial variation of the centroid energy in 3C~397 is of particular interest. 3C~397 was initially theorized to have a Type Ia progenitor \citep{Chen1999}. Subsequently a CC progenitor was proposed due to the abundance pattern observed with \textit{ASCA} and its proximity to a molecular cloud \citep{Safi-Harb2000}. Y14 concluded that the remnant's centroid energy could be consistent with both interpretations, based on the Fe K$\alpha$ line energy, but favored the Type Ia hypothesis based on their simulations. Most recently, \cite{Yamaguchi2015} and \cite{Martinez-Rodriguez2020} advocated for a Type~Ia origin using the observed enrichment of Mn and Ni from \textit{Suzaku} and a comparison of the line energies and fluxes with hydrodynamic models, respectively. 

Y14, \cite{Yamaguchi2015}, and \cite{Martinez-Rodriguez2020} all fit the Fe K$\alpha$ centroid energy using the emission from the entire remnant. This inherently assumes that the overall Fe K$\alpha$ centroid energy would be similar to the energies extracted by fitting different regions in the remnant, which is true if the plasma is relatively homogeneous. However, in the case of 3C~397, the Fe K$\alpha$ centroid energies from some regions are more consistent with a CC origin. In order for the dividing line proposed by Y14 to be consistent with a CC designation for 3C~397, it could be revised to 6500~eV, which would place all the  centroid energies in the CC region, consistent with \citet{Safi-Harb2000, SafiHarb2005}. This would, however, contradict other indicators, such as the enhanced abundance of Mn and Ni, which  suggest a Type Ia progenitor \citep{Yamaguchi2015}. If the remnant were instead definitively classified as Type Ia, then the dividing line would need to be moved up considerably, to $\gtrapprox$ 6600~eV; however, this would contradict the classification of G292.0+1.8 and G350.1-0.3 as CC remnants. 

These results show that using only the Fe K$\alpha$ centroid energy as a discriminant between Type Ia and core-collapse progenitors is not always a robust technique. We therefore advocate caution when using the centroid alone to type remnants, particularly in cases where the remnant is not spatially resolved.

\section{Conclusions}
\label{sec:concl}
In this paper, we investigate Fe K$\alpha$ centroids in 6 SNRs. We detect the Fe K$\alpha$ line in 4 of the remnants and are able to conduct spatially resolved fits for N132D, W49B, and 3C~397. The centroid energies derived for the entire emitting regions of N132D, W49B, and 3C~397 are consistent with Y14. For DEM L71, where the Fe K$\alpha$ line was not detected by Y14, our observed value falls within the Type Ia parameter space reported by Y14; this agrees with prior studies on the origin of DEM L71.

Our main conclusions are:

\begin{enumerate}
    \item The dividing line between Fe K$\alpha$ centroid energies in CC and Type Ia remnants, taken by Y14 to lie at 6550 eV, is not well-defined and should not be used as the sole discriminator of type. Numerical models \citep{patnaudeetal15,Martinez-Rodriguez2018,Jacovich2021} appear to validate this conclusion.
    \item Prior studies, including Y14, extracted the spectrum from the entire remnant, but the centroid energies extracted from different regions may not agree with the centroid from the entire remnant.  Our spatially resolved fits in 3C~397 reveal statistically significant variations in the Fe K$\alpha$ centroid energy up to 100~eV,  spanning both the Type Ia and CC regions designated by Y14; for G344.7-0.1 the variations reached 130 eV \citep{Fukushima2020}.
    \item The discrepancy between the Fe K$\alpha$ centroid energies of 3C~397, its classification by various means, and the dividing line suggested by Y14, cannot be resolved by simply revising the position of the dividing line. Although the agreement could be improved by shifting the dividing line such that all the centroid energies for 3C~397 lie on the CC side, this would contradict the remnant type obtained by other means. Shifting the dividing line so that all 3C~397 energies lie on the Type Ia side would contradict the types of other known CC remnants.
    \item Even if the centroids obtained from various regions are consistent with that obtained from the entire remnant, they may not agree with other diagnostics of the remnant type. The Fe K$\alpha$ centroids in various parts of W49B all appear to have similar energies and agree with the centroid energy measured from the entire remnant. These high centroid energies however are not consistent with recent abundance studies, and comparison to theoretical models, which suggested a Type Ia progenitor \citep{zv18, siegeletal20b}. These studies contradicted prior ones that favored a core-collapse origin by comparing inferred abundances with a limited set of theoretical models \citep{Lopez2009, Lopez2013}. A study by \cite{Sun2020}, which was inconclusive on the progenitor type, even conceded that the large Mn/Fe ratio could not be produced by any core-collapse model, thus hinting at a Type Ia origin.
    \item The centroid energy in different regions of a remnant will be affected by the density of the surrounding material into which the remnant is expanding. In 3C~397, the regions with the highest Fe K$\alpha$ centroid energy are coincident with the regions of remnant interaction with a high density molecular cloud, whereas regions with lower Fe K$\alpha$ centroid energy are expanding into lower densities. While spherically symmetric models have generally confirmed the correlation between centroid energy and ambient density \citep{patnaudeetal15,Martinez-Rodriguez2018,Jacovich2021}, they are unable to model the variations of the centroid energy within a remnant due to an inhomogeneous surrounding medium.
\end{enumerate}

Our study does not necessarily indicate or revise the classification of 3C~397. Our intention is to show that the energy of the Fe K$\alpha$ centroid in different parts of the remnant varies based on the density into which the remnant is expanding, and may be inconsistent with the centroid energy of the entire remnant. Large density variations around the remnant could result in large variations in the centroid energy. Thus far such variations have been observed in only a few remnants. There exists a clear need for more studies of the energy of the Fe K$\alpha$ centroid in different regions of large, spatially-resolved remnants.

Our results thus demonstrate the intricacies involved when using  the centroid energy of the Fe K$\alpha$ line as the sole discriminant to type SNRs. We therefore urge caution when using the overall Fe K$\alpha$ centroid energy as a diagnostic of remnant type, especially in remnants where the density of the surrounding medium varies considerably over the remnant's surface, or the level of such spatial variations is unknown. Furthermore, the dividing line between CC and Type Ia remnants in terms of centroid energy is somewhat arbitrary and subject to revision, so care should be taken when using the centroid energy to type remnants close to the line. Although the Fe K$\alpha$ centroid energy remains a suitable technique for simpler remnants and may be applicable in many cases, complexities in the ambient medium can shift the centroid energies and hinder classification attempts. Therefore the centroid energy by itself is not a sufficient diagnostic when attempting to find a SNR's type. While considering the line luminosity alongside the centroid energy may make the overall combination a better discriminant, this needs to be clearly demonstrated by observational studies. Unfortunately, luminosity depends upon knowing the accurate distance to the remnant, which is not always possible for Galactic remnants, and adds one more variable to the puzzle.

\ \\
{\bf Acknowledgments:}
We thank the referee for comments that have helped to
substantially improve the paper. We also thank Franz Bauer for comments regarding the Fe K$\alpha$ line. This work was partially supported by NASA ADAP grant NNX15AF03G to Pennsylvania State University, with subcontracts to the University of Chicago and  Northwestern  University.  VVD is supported by NSF grant 1911061. Based on observations obtained with \textit{XMM-Newton}, an ESA science mission with instruments and contributions directly funded by ESA Member States and NASA.
\software{ \texttt{XSPEC} \citep{XSPEC}, \texttt{Matplotlib} \citep{matplotlib},
\texttt{numpy} \citep{numpy}, \texttt{scipy} \citep{2020SciPy-NMeth}}

\facilities{XMM(EPIC)}

\bibliography{paper}%

\begin{thebibliography}{}
\expandafter\ifx\csname natexlab\endcsname\relax\def\natexlab#1{#1}\fi
\providecommand{\url}[1]{\href{#1}{#1}}
\providecommand{\dodoi}[1]{doi:~\href{http://doi.org/#1}{\nolinkurl{#1}}}
\providecommand{\doeprint}[1]{\href{http://ascl.net/#1}{\nolinkurl{http://ascl.net/#1}}}
\providecommand{\doarXiv}[1]{\href{https://arxiv.org/abs/#1}{\nolinkurl{https://arxiv.org/abs/#1}}}

\bibitem[{{Alan} {et~al.}(2019){Alan}, {Park}, \& {Bilir}}]{Neslihan2019}
{Alan}, N., {Park}, S., \& {Bilir}, S. 2019, \apj, 873, 53,
  \dodoi{10.3847/1538-4357/aaf882}

\bibitem[{{Aldering} {et~al.}(2006){Aldering}, {Antilogus}, {Bailey}, {Baltay},
  {Bauer}, {Blanc}, {Bongard}, {Copin}, {Gangler}, {Gilles}, {Kessler},
  {Kocevski}, {Lee}, {Loken}, {Nugent}, {Pain}, {P{\'e}contal}, {Pereira},
  {Perlmutter}, {Rabinowitz}, {Rigaudier}, {Scalzo}, {Smadja}, {Thomas},
  {Wang}, {Weaver}, \& {Nearby Supernova Factory}}]{alderingetal06}
{Aldering}, G., {Antilogus}, P., {Bailey}, S., {et~al.} 2006, \apj, 650, 510,
  \dodoi{10.1086/507020}

\bibitem[{{Arnaud}(1996)}]{XSPEC}
{Arnaud}, K.~A. 1996, in Astronomical Society of the Pacific Conference Series,
  Vol. 101, Astronomical Data Analysis Software and Systems V, ed. G.~H.
  {Jacoby} \& J.~{Barnes}, 17

\bibitem[{{Bevington}(1969)}]{Bevington1969}
{Bevington}, P.~R. 1969, {Data reduction and error analysis for the physical
  sciences}

\bibitem[{{Bhalerao} {et~al.}(2019){Bhalerao}, {Park}, {Schenck}, {Post}, \&
  {Hughes}}]{bhaleraoetal19}
{Bhalerao}, J., {Park}, S., {Schenck}, A., {Post}, S., \& {Hughes}, J.~P. 2019,
  \apj, 872, 31, \dodoi{10.3847/1538-4357/aafafd}

\bibitem[{{Bochenek} {et~al.}(2018){Bochenek}, {Dwarkadas}, {Silverman}, {Fox},
  {Chevalier}, {Smith}, \& {Filippenko}}]{Bochenek2018}
{Bochenek}, C.~D., {Dwarkadas}, V.~V., {Silverman}, J.~M., {et~al.} 2018,
  \mnras, 473, 336, \dodoi{10.1093/mnras/stx2029}

\bibitem[{{Chen} {et~al.}(1999){Chen}, {Sun}, {Wang}, \& {Yin}}]{Chen1999}
{Chen}, Y., {Sun}, M., {Wang}, Z.-R., \& {Yin}, Q.~F. 1999, \apj, 520, 737,
  \dodoi{10.1086/307489}

\bibitem[{{Chevalier} \& {Liang}(1989)}]{cl89}
{Chevalier}, R.~A., \& {Liang}, E.~P. 1989, \apj, 344, 332,
  \dodoi{10.1086/167802}

\bibitem[{{Chiotellis} {et~al.}(2020){Chiotellis}, {Boumis}, \&
  {Spetsieri}}]{Chiotellis2020}
{Chiotellis}, A., {Boumis}, P., \& {Spetsieri}, Z.~T. 2020, Galaxies, 8, 38,
  \dodoi{10.3390/galaxies8020038}

\bibitem[{{Chiotellis} {et~al.}(2012){Chiotellis}, {Schure}, \&
  {Vink}}]{Chiotellis2012}
{Chiotellis}, A., {Schure}, K.~M., \& {Vink}, J. 2012, \aap, 537, A139,
  \dodoi{10.1051/0004-6361/201014754}

\bibitem[{{Dwarkadas}(2005)}]{Dwarkadas2005}
{Dwarkadas}, V.~V. 2005, \apj, 630, 892, \dodoi{10.1086/432109}

\bibitem[{{Dwarkadas}(2007)}]{dwarkadas07}
---. 2007, \apj, 667, 226, \dodoi{10.1086/520670}

\bibitem[{{Eriksen} {et~al.}(2001){Eriksen}, {Morse}, {Kirshner}, \&
  {Winkler}}]{Eriksen2001}
{Eriksen}, K.~A., {Morse}, J.~A., {Kirshner}, R.~P., \& {Winkler}, P.~F. 2001,
  in American Institute of Physics Conference Series, Vol. 565, Young Supernova
  Remnants, ed. S.~S. {Holt} \& U.~{Hwang}, 193--196, \dodoi{10.1063/1.1377093}

\bibitem[{{Frank} {et~al.}(2019){Frank}, {Dwarkadas}, {Panfichi}, {Crum}, \&
  {Burrows}}]{franketal19}
{Frank}, K.~A., {Dwarkadas}, V., {Panfichi}, A., {Crum}, R.~M., \& {Burrows},
  D.~N. 2019, \apj, 875, 14, \dodoi{10.3847/1538-4357/ab0e81}

\bibitem[{{Fukushima} {et~al.}(2020){Fukushima}, {Yamaguchi}, {Slane}, {Park},
  {Katsuda}, {Sano}, {Lopez}, {Plucinsky}, {Kobayashi}, \&
  {Matsushita}}]{Fukushima2020}
{Fukushima}, K., {Yamaguchi}, H., {Slane}, P.~O., {et~al.} 2020, \apj, 897, 62,
  \dodoi{10.3847/1538-4357/ab94a6}

\bibitem[{{Ghavamian} {et~al.}(2003){Ghavamian}, {Rakowski}, {Hughes}, \&
  {Williams}}]{Ghavamian2003}
{Ghavamian}, P., {Rakowski}, C.~E., {Hughes}, J.~P., \& {Williams}, T.~B. 2003,
  \apj, 590, 833, \dodoi{10.1086/375161}

\bibitem[{{Hughes} {et~al.}(2003){Hughes}, {Ghavamian}, {Rakowski}, \&
  {Slane}}]{hughesetal03}
{Hughes}, J.~P., {Ghavamian}, P., {Rakowski}, C.~E., \& {Slane}, P.~O. 2003,
  \apjl, 582, L95, \dodoi{10.1086/367760}

\bibitem[{{Hughes} {et~al.}(2000){Hughes}, {Rakowski}, \&
  {Decourchelle}}]{Hughes2000}
{Hughes}, J.~P., {Rakowski}, C.~E., \& {Decourchelle}, A. 2000, \apjl, 543,
  L61, \dodoi{10.1086/312945}

\bibitem[{{Hughes} {et~al.}(1995){Hughes}, {Hayashi}, {Helfand}, {Hwang},
  {Itoh}, {Kirshner}, {Koyama}, {Markert}, {Tsunemi}, \& {Woo}}]{Hughes1995}
{Hughes}, J.~P., {Hayashi}, I., {Helfand}, D., {et~al.} 1995, \apjl, 444, L81,
  \dodoi{10.1086/187865}

\bibitem[{{Hunter}(2007)}]{matplotlib}
{Hunter}, J.~D. 2007, Computing in Science Engineering, 9, 90

\bibitem[{{Jacovich} {et~al.}(2021){Jacovich}, {Patnaude}, {Slane}, {Badenes},
  {Lee}, {Nagataki}, \& {Milisavljevic}}]{Jacovich2021}
{Jacovich}, T., {Patnaude}, D., {Slane}, P., {et~al.} 2021, arXiv e-prints,
  arXiv:2103.07980.
\newblock \doarXiv{2103.07980}

\bibitem[{{Kilpatrick} {et~al.}(2016){Kilpatrick}, {Bieging}, \&
  {Rieke}}]{Kilpatrick2016}
{Kilpatrick}, C.~D., {Bieging}, J.~H., \& {Rieke}, G.~H. 2016, \apj, 816, 1,
  \dodoi{10.3847/0004-637X/816/1/1}

\bibitem[{{Kraft} {et~al.}(1991){Kraft}, {Burrows}, \& {Nousek}}]{Kraft1991}
{Kraft}, R.~P., {Burrows}, D.~N., \& {Nousek}, J.~A. 1991, \apj, 374, 344,
  \dodoi{10.1086/170124}

\bibitem[{{Kumar} {et~al.}(2014){Kumar}, {Safi-Harb}, {Slane}, \&
  {Gotthelf}}]{Kumar2014}
{Kumar}, H.~S., {Safi-Harb}, S., {Slane}, P.~O., \& {Gotthelf}, E.~V. 2014,
  \apj, 781, 41, \dodoi{10.1088/0004-637X/781/1/41}

\bibitem[{{Leahy} \& {Ranasinghe}(2016)}]{Leahy2016}
{Leahy}, D.~A., \& {Ranasinghe}, S. 2016, \apj, 817, 74,
  \dodoi{10.3847/0004-637X/817/1/74}

\bibitem[{{Lopez} {et~al.}(2013){Lopez}, {Pearson}, {Ramirez-Ruiz}, {Castro},
  {Yamaguchi}, {Slane}, \& {Smith}}]{Lopez2013}
{Lopez}, L.~A., {Pearson}, S., {Ramirez-Ruiz}, E., {et~al.} 2013, \apj, 777,
  145, \dodoi{10.1088/0004-637X/777/2/145}

\bibitem[{{Lopez} {et~al.}(2011){Lopez}, {Ramirez-Ruiz}, {Huppenkothen},
  {Badenes}, \& {Pooley}}]{Lopez2011}
{Lopez}, L.~A., {Ramirez-Ruiz}, E., {Huppenkothen}, D., {Badenes}, C., \&
  {Pooley}, D.~A. 2011, \apj, 732, 114, \dodoi{10.1088/0004-637X/732/2/114}

\bibitem[{{Lopez} {et~al.}(2009){Lopez}, {Ramirez-Ruiz}, {Pooley}, \&
  {Jeltema}}]{Lopez2009}
{Lopez}, L.~A., {Ramirez-Ruiz}, E., {Pooley}, D.~A., \& {Jeltema}, T.~E. 2009,
  \apj, 691, 875, \dodoi{10.1088/0004-637X/691/1/875}

\bibitem[{{Mart{\'\i}nez-Rodr{\'\i}guez}
  {et~al.}(2018){Mart{\'\i}nez-Rodr{\'\i}guez}, {Badenes}, {Lee}, {Patnaude},
  {Foster}, {Yamaguchi}, {Auchettl}, {Bravo}, {Slane}, {Piro}, {Park}, \&
  {Nagataki}}]{Martinez-Rodriguez2018}
{Mart{\'\i}nez-Rodr{\'\i}guez}, H., {Badenes}, C., {Lee}, S.-H., {et~al.} 2018,
  \apj, 865, 151, \dodoi{10.3847/1538-4357/aadaec}

\bibitem[{{Mart{\'\i}nez-Rodr{\'\i}guez}
  {et~al.}(2020){Mart{\'\i}nez-Rodr{\'\i}guez}, {Lopez}, {Auchettl}, {Badenes},
  {Holland-Ashford}, {Patnaude}, {Lee}, {Foster}, \&
  {Slane}}]{Martinez-Rodriguez2020}
{Mart{\'\i}nez-Rodr{\'\i}guez}, H., {Lopez}, L.~A., {Auchettl}, K., {et~al.}
  2020, arXiv e-prints, arXiv:2006.08681.
\newblock \doarXiv{2006.08681}

\bibitem[{{Morse} {et~al.}(1995){Morse}, {Winkler}, \& {Kirshner}}]{Morse1995}
{Morse}, J.~A., {Winkler}, P.~F., \& {Kirshner}, R.~P. 1995, \aj, 109, 2104,
  \dodoi{10.1086/117436}

\bibitem[{{Ozawa} {et~al.}(2009){Ozawa}, {Koyama}, {Yamaguchi}, {Masai}, \&
  {Tamagawa}}]{ozawaetal09}
{Ozawa}, M., {Koyama}, K., {Yamaguchi}, H., {Masai}, K., \& {Tamagawa}, T.
  2009, \apjl, 706, L71, \dodoi{10.1088/0004-637X/706/1/L71}

\bibitem[{{Patnaude} {et~al.}(2012){Patnaude}, {Badenes}, {Park}, \&
  {Laming}}]{Patnaude2012}
{Patnaude}, D.~J., {Badenes}, C., {Park}, S., \& {Laming}, J.~M. 2012, \apj,
  756, 6, \dodoi{10.1088/0004-637X/756/1/6}

\bibitem[{{Patnaude} {et~al.}(2015){Patnaude}, {Lee}, {Slane}, {Badenes},
  {Heger}, {Ellison}, \& {Nagataki}}]{patnaudeetal15}
{Patnaude}, D.~J., {Lee}, S.-H., {Slane}, P.~O., {et~al.} 2015, \apj, 803, 101,
  \dodoi{10.1088/0004-637X/803/2/101}

\bibitem[{{Protassov} {et~al.}(2002){Protassov}, {van Dyk}, {Connors},
  {Kashyap}, \& {Siemiginowska}}]{Protassov2002}
{Protassov}, R., {van Dyk}, D.~A., {Connors}, A., {Kashyap}, V.~L., \&
  {Siemiginowska}, A. 2002, \apj, 571, 545, \dodoi{10.1086/339856}

\bibitem[{{Pye} {et~al.}(1984){Pye}, {Becker}, {Seward}, \& {Thomas}}]{pye1984}
{Pye}, J.~P., {Becker}, R.~H., {Seward}, F.~D., \& {Thomas}, N. 1984, \mnras,
  207, 649, \dodoi{10.1093/mnras/207.3.649}

\bibitem[{{Quirola-V{\'a}squez} {et~al.}(2019){Quirola-V{\'a}squez}, {Bauer},
  {Dwarkadas}, {Badenes}, {Brandt}, {Nymark}, \&
  {Walton}}]{quirolavasquezetal19}
{Quirola-V{\'a}squez}, J., {Bauer}, F.~E., {Dwarkadas}, V.~V., {et~al.} 2019,
  \mnras, 490, 4536, \dodoi{10.1093/mnras/stz2858}

\bibitem[{{Rest} {et~al.}(2005){Rest}, {Suntzeff}, {Olsen}, {Prieto}, {Smith},
  {Welch}, {Becker}, {Bergmann}, {Clocchiatti}, {Cook}, {Garg}, {Huber},
  {Miknaitis}, {Minniti}, {Nikolaev}, \& {Stubbs}}]{Rest2005}
{Rest}, A., {Suntzeff}, N.~B., {Olsen}, K., {et~al.} 2005, \nat, 438, 1132,
  \dodoi{10.1038/nature04365}

\bibitem[{{Safi-Harb} {et~al.}(2005){Safi-Harb}, {Dubner}, {Petre}, {Holt}, \&
  {Durouchoux}}]{SafiHarb2005}
{Safi-Harb}, S., {Dubner}, G., {Petre}, R., {Holt}, S.~S., \& {Durouchoux}, P.
  2005, \apj, 618, 321, \dodoi{10.1086/425960}

\bibitem[{{Safi-Harb} {et~al.}(2000){Safi-Harb}, {Petre}, {Arnaud}, {Keohane},
  {Borkowski}, {Dyer}, {Reynolds}, \& {Hughes}}]{Safi-Harb2000}
{Safi-Harb}, S., {Petre}, R., {Arnaud}, K.~A., {et~al.} 2000, \apj, 545, 922,
  \dodoi{10.1086/317823}

\bibitem[{{Sawada} {et~al.}(2019){Sawada}, {Tachibana}, {Uchida}, {Ito},
  {Matsumura}, {Bamba}, {Tsuru}, \& {Tanaka}}]{sawadaetal19}
{Sawada}, M., {Tachibana}, K., {Uchida}, H., {et~al.} 2019, \pasj, 71, 61,
  \dodoi{10.1093/pasj/psz036}

\bibitem[{{Sezer} {et~al.}(2018){Sezer}, {Ergin}, {Yamazaki}, {Ohira}, \&
  {Cesur}}]{sezeretal18}
{Sezer}, A., {Ergin}, T., {Yamazaki}, R., {Ohira}, Y., \& {Cesur}, N. 2018,
  \mnras, 481, 1416, \dodoi{10.1093/mnras/sty2387}

\bibitem[{{Sharda} {et~al.}(2020){Sharda}, {Gaetz}, {Kashyap}, \&
  {Plucinsky}}]{Sharda2020}
{Sharda}, P., {Gaetz}, T.~J., {Kashyap}, V.~L., \& {Plucinsky}, P.~P. 2020,
  \apj, 894, 145, \dodoi{10.3847/1538-4357/ab8a46}

\bibitem[{{Siegel} {et~al.}(2020{\natexlab{a}}){Siegel}, {Dwarkadas}, {Frank},
  {Burrows}, \& {Panfichi}}]{Siegel2020}
{Siegel}, J., {Dwarkadas}, V.~V., {Frank}, K., {Burrows}, D.~N., \& {Panfichi},
  A. 2020{\natexlab{a}}, Astronomische Nachrichten, 341, 163,
  \dodoi{10.1002/asna.202023773}

\bibitem[{{Siegel} {et~al.}(2020{\natexlab{b}}){Siegel}, {Dwarkadas}, {Frank},
  \& {Burrows}}]{siegeletal20b}
{Siegel}, J., {Dwarkadas}, V.~V., {Frank}, K.~A., \& {Burrows}, D.~N.
  2020{\natexlab{b}}, \apj, 904, 175, \dodoi{10.3847/1538-4357/abbfa9}

\bibitem[{{Silverman} {et~al.}(2013){Silverman}, {Nugent}, {Gal-Yam},
  {Sullivan}, {Howell}, {Filippenko}, {Arcavi}, {Ben-Ami}, {Bloom}, {Cenko},
  {Cao}, {Chornock}, {Clubb}, {Coil}, {Foley}, {Graham}, {Griffith}, {Horesh},
  {Kasliwal}, {Kulkarni}, {Leonard}, {Li}, {Matheson}, {Miller}, {Modjaz},
  {Ofek}, {Pan}, {Perley}, {Poznanski}, {Quimby}, {Steele}, {Sternberg}, {Xu},
  \& {Yaron}}]{silvermanetal13}
{Silverman}, J.~M., {Nugent}, P.~E., {Gal-Yam}, A., {et~al.} 2013, \apjs, 207,
  3, \dodoi{10.1088/0067-0049/207/1/3}

\bibitem[{{Sun} \& {Chen}(2020)}]{Sun2020}
{Sun}, L., \& {Chen}, Y. 2020, \apj, 893, 90, \dodoi{10.3847/1538-4357/ab8001}

\bibitem[{{Tenorio-Tagle} {et~al.}(1990){Tenorio-Tagle}, {Bodenheimer},
  {Franco}, \& {Rozyczka}}]{tbfr90}
{Tenorio-Tagle}, G., {Bodenheimer}, P., {Franco}, J., \& {Rozyczka}, M. 1990,
  \mnras, 244, 563

\bibitem[{{Tenorio-Tagle} {et~al.}(1991){Tenorio-Tagle}, {Rozyczka}, {Franco},
  \& {Bodenheimer}}]{trfb91}
{Tenorio-Tagle}, G., {Rozyczka}, M., {Franco}, J., \& {Bodenheimer}, P. 1991,
  \mnras, 251, 318, \dodoi{10.1093/mnras/251.2.318}

\bibitem[{{Tian} \& {Leahy}(2008)}]{Tian2008}
{Tian}, W.~W., \& {Leahy}, D.~A. 2008, \apj, 677, 292, \dodoi{10.1086/529120}

\bibitem[{{van der Heyden} {et~al.}(2002){van der Heyden}, {Behar}, {Vink},
  {Rasmussen}, {Kaastra}, {Bleeker}, {Kahn}, \& {Mewe}}]{vanderHeyden2002}
{van der Heyden}, K.~J., {Behar}, E., {Vink}, J., {et~al.} 2002, \aap, 392,
  955, \dodoi{10.1051/0004-6361:20020963}

\bibitem[{{van der Walt} {et~al.}(2011){van der Walt}, {Colbert}, \&
  {Varoquaux}}]{numpy}
{van der Walt}, S., {Colbert}, S.~C., \& {Varoquaux}, G. 2011, Computing in
  Science and Engineering, 13, 22, \dodoi{10.1109/MCSE.2011.37}

\bibitem[{{Vink}(2012)}]{Vink2012}
{Vink}, J. 2012, \aapr, 20, 49, \dodoi{10.1007/s00159-011-0049-1}

\bibitem[{Virtanen {et~al.}(2020)Virtanen, Gommers, Oliphant, Haberland, Reddy,
  Cournapeau, Burovski, Peterson, Weckesser, Bright, {van der Walt}, Brett,
  Wilson, Millman, Mayorov, Nelson, Jones, Kern, Larson, Carey, Polat, Feng,
  Moore, {VanderPlas}, Laxalde, Perktold, Cimrman, Henriksen, Quintero, Harris,
  Archibald, Ribeiro, Pedregosa, {van Mulbregt}, \& {SciPy 1.0
  Contributors}}]{2020SciPy-NMeth}
Virtanen, P., Gommers, R., Oliphant, T.~E., {et~al.} 2020, Nature Methods, 17,
  261, \dodoi{10.1038/s41592-019-0686-2}

\bibitem[{{Williams} {et~al.}(2011){Williams}, {Blair}, {Blondin}, {Borkowski},
  {Ghavamian}, {Long}, {Raymond}, {Reynolds}, {Rho}, \&
  {Winkler}}]{williamsetal11}
{Williams}, B.~J., {Blair}, W.~P., {Blondin}, J.~M., {et~al.} 2011, \apj, 741,
  96, \dodoi{10.1088/0004-637X/741/2/96}

\bibitem[{{Yamaguchi} {et~al.}(2021){Yamaguchi}, {Acero}, {Li}, \&
  {Chu}}]{yamaguchietal21}
{Yamaguchi}, H., {Acero}, F., {Li}, C.-J., \& {Chu}, Y.-H. 2021, \apjl, 910,
  L24, \dodoi{10.3847/2041-8213/abee8a}

\bibitem[{{Yamaguchi} {et~al.}(2014){Yamaguchi}, {Badenes}, {Petre}, {Nakano},
  {Castro}, {Enoto}, {Hiraga}, {Hughes}, {Maeda}, {Nobukawa}, {Safi-Harb},
  {Slane}, {Smith}, \& {Uchida}}]{yamaguchi2014}
{Yamaguchi}, H., {Badenes}, C., {Petre}, R., {et~al.} 2014, \apjl, 785, L27,
  \dodoi{10.1088/2041-8205/785/2/L27}

\bibitem[{{Yamaguchi} {et~al.}(2015){Yamaguchi}, {Badenes}, {Foster}, {Bravo},
  {Williams}, {Maeda}, {Nobukawa}, {Eriksen}, {Brickhouse}, {Petre}, \&
  {Koyama}}]{Yamaguchi2015}
{Yamaguchi}, H., {Badenes}, C., {Foster}, A.~R., {et~al.} 2015, \apjl, 801,
  L31, \dodoi{10.1088/2041-8205/801/2/L31}

\bibitem[{{Zhou} {et~al.}(2021){Zhou}, {Leung}, {Li}, {Nomoto}, {Vink}, \&
  {Chen}}]{Zhou2021}
{Zhou}, P., {Leung}, S.-C., {Li}, Z., {et~al.} 2021, \apj, 908, 31,
  \dodoi{10.3847/1538-4357/abbd45}

\bibitem[{{Zhou} \& {Vink}(2018)}]{zv18}
{Zhou}, P., \& {Vink}, J. 2018, \aap, 615, A150,
  \dodoi{10.1051/0004-6361/201731583}

\bibitem[{{Zhu} {et~al.}(2014){Zhu}, {Tian}, \& {Zuo}}]{Zhu2014}
{Zhu}, H., {Tian}, W.~W., \& {Zuo}, P. 2014, \apj, 793, 95,
  \dodoi{10.1088/0004-637X/793/2/95}

\end{thebibliography}

\end{document}